\documentclass[10pt, conference]{IEEEtran}
\IEEEoverridecommandlockouts
\usepackage{cite}
\usepackage{amsmath,amssymb,amsfonts}
\usepackage{algorithmic}
\usepackage{graphicx}
\usepackage{textcomp}
\usepackage{xcolor}
\usepackage{url}
\usepackage{hyperref}
\usepackage{cleveref}

\def\BibTeX{{\rm B\kern-.05em{\sc i\kern-.025em b}\kern-.08em
    T\kern-.1667em\lower.7ex\hbox{E}\kern-.125emX}}
\begin{document}

\title{A Photonic Physically Unclonable Function's Resilience to Multiple-Valued Machine Learning Attacks\\
\thanks{Funded by Anametric, Inc.}
}

\author{
\IEEEauthorblockN{Jessie M. Henderson\textsuperscript{1},
Elena R. Henderson\textsuperscript{1},
Clayton A. Harper\textsuperscript{1},
Hiva Shahoei\textsuperscript{1}, \\
William V. Oxford\textsuperscript{2},
Eric C. Larson\textsuperscript{1},
Duncan L. MacFarlane\textsuperscript{1}, and
Mitchell A. Thornton\textsuperscript{1}}
\IEEEauthorblockA{\textsuperscript{1}Darwin Deason Institute for Cyber Security, Southern Methodist University, Dallas, TX, USA}
\IEEEauthorblockA{\textsuperscript{2}Anametric, Inc., Austin, TX, USA}

\IEEEauthorblockA{\textsuperscript{1}\{hendersonj, erhenderson, caharper, hshahoei, eclarson, dmacfarlane, mitch\}@smu.edu}

\IEEEauthorblockA{\textsuperscript{2}oxford@anametric.com}
}

\maketitle

\begin{abstract}
Physically unclonable functions (PUFs) identify integrated circuits using nonlinearly-related challenge-response pairs (CRPs).
Ideally, the relationship between challenges and corresponding responses is unpredictable, even if a subset of CRPs is known.
Previous work developed a photonic PUF offering improved security compared to non-optical counterparts.
Here, we investigate this PUF's susceptibility to Multiple-Valued-Logic-based machine learning attacks.
We find that approximately $1$,$000$ CRPs are necessary to train models that predict response bits better than random chance.
Given the significant challenge of acquiring a vast number of CRPs from a photonic PUF, our results demonstrate photonic PUF resilience against such attacks.
\end{abstract}


\section{Introduction}
Physically unclonable functions (PUFs) provide identifying signatures for integrated circuits (ICs) by responding to challenges in a way that is difficult to predict but straightforward to verify.
The theory and implementation of conventional PUFs are well-understood, lending several architectures that depend upon probabilistic geometric anomalies introduced during circuit fabrication \cite{guajardo07,maes13,maiti11,mcgrath19,pelgrom89}.
Such PUFs are widely applicable in situations such as IC authentication, trusted system development, and hardware-level cybersecurity \cite{lofstrom00,pappu02,gassend02,gassend02a,lee04}.
Unfortunately, several PUF implementations have also proven vulnerable to malicious exploits, especially those utilizing machine learning (ML).
Such attacks seek to predict responses for as-of-yet unseen challenges by learning from a set of known challenge-response pairs (CRPs) \cite{delvaux19,hazari21,kumar18,ge20,santikellur19,machida15,aghaie21,khalafalla19,saha13,laguduva20,guo16,ruhrmair10,van13,yoon20}, and they can be effective even when no mathematical representation of a PUF's structure is known \cite{ganji16}.
Consequently, it is worth exploring alternative PUF structures, of which photonic PUFs are a promising candidate.

Photonic PUFs leverage sensitive manufacturing tolerances of photonic integrated circuit (PIC) components to provide CRPs that are difficult to predict, in part because accurately representing the function that maps challenges to responses may require infinitely many matrix-vector calculations \cite{macfarlane23,jones42}.
Additionally, collecting CRPs with which to train ML-based PUF attacks is more challenging for photonic PUFs than for non-optical variants, because the use of light as input and output makes photonic PUFs less prone to physical and electromagnetic interference attacks\cite{helfmeier14}.
MacFarlane et al. experimentally demonstrated that the outputs of a photonic PUF can be modeled using Multiple-Valued Logic (MVL)\cite{macfarlane23}.
In this work, we analyze that photonic PUF's susceptibility to ML-based attacks such as those applied to conventional IC PUFs.
We consider network configurations utilizing different MVL digit representations for network inputs and outputs.
Applying these different radix systems may be a useful feature engineering technique, facilitating a potentially smoother optimization space that the network can exploit for improved predictive performance, or alternatively, a representation that is more difficult to exploit for PUF-based attacks.
Therefore, MVL representations of CRPs may enhance network performance, motivating us to assess the photonic PUF's susceptibility to these differing attack strategies.

The remainder of the paper proceeds as follows: in \Cref{sec:background}, we briefly contextualize the photonic PUF and review ML attacks on conventional PUFs.
\Cref{sec:attacking_the_puf} describes the neural network attack, including a computational PUF model developed to generate sufficient data for training the models.
In \Cref{sec:results_and_discussion}, we compare the results of attacking the PUF with the binary- and MVL-based networks, with a particular focus on the number of CRPs required to predict responses more accurately than chance.
We conclude by contextualizing our findings and briefly considering avenues for further research.

\section{Background}\label{sec:background}

\subsection{A Photonic PUF}
\begin{figure}[tbh]
\centerline{\includegraphics[width=0.85\linewidth]{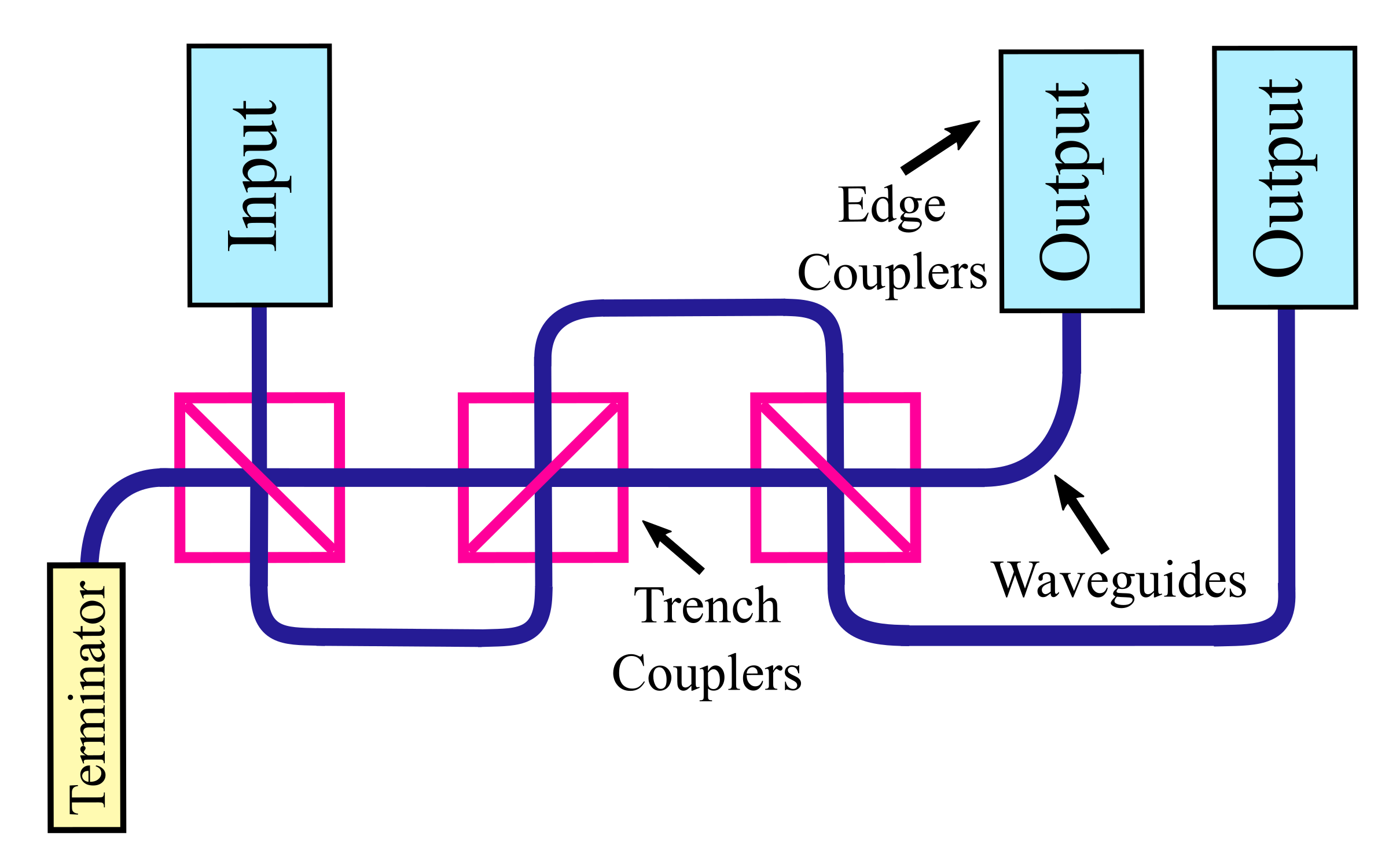}}
\caption{Overview of the PIC PUF layout. Note that the PUF described in this work uses only one of the two available outputs to generate CRPs.}
\label{fig:photonic_puf_layout}
\end{figure}

Photonic PUF design is a nascent area of research, because the underlying physics and implementation mechanisms differ dramatically from those of mature electronic components and fabrication techniques.
MacFarlane et al. developed a PUF using a photonic integrated chip (PIC) layout with trench couplers whose manufacturing sensitivites and feature sizes of nanoscale structures are studied in Refs. \cite{shahoei22} and \cite{macfarlane23}.
\Cref{fig:photonic_puf_layout} illustrates the structure of the PUF (three trench couplers connected via waveguides and terminated with edge couplers).
Theoretically, the input state of polarization (SOP) of an optical signal is highly dependent upon miniscule perturbations in PIC component geometry, including those that derive from stresses and strains imposed upon the PIC during fabrication and packaging.
Experimental results from Shahoei et al. and MacFarlane et al. illustrate this desirably-strong dependence on precise geometry \cite{shahoei22,macfarlane23}.

As analysis of the PIC PUF design is provided in previous work, we make solely two notes here.
First, the optical components of the PIC PUF can be mathematically represented using matrices and Jones calculus \cite{jones42}.
However, due to the variations in component geometry, an accurate representation requires potentially infinitely many matrices.
This makes transfer function replication intractably difficult, and motivates our study of ML-based PUF attacks that learn such relationships without explicitly replicating the mapping function.
Second, training ML models to attack PUFs often requires substantial (\textit{e.g.}, thousands to millions) CRP training observations from which to learn patterns.
Photonic PUFs are superior to conventional IC variants in that efficiently obtaining large numbers of CRPs is currently impractical; indeed, this is why we sought to learn the number of CRPs required to train models that predict responses better than chance.
While difficulty of obtaining CRPs is thus ordinarily a PUF strength \cite{ruhrmair10,ruhrmair14}, it is a challenge in the context of assessing the PUF’s resilience to ML-based attacks.
We thus adopted a strategy similar to that of Santikellur et al. by developing a computational model (written in the \texttt{Python} programming language) to efficiently simulate PUFs which can then be analyzed for susceptibility to ML-based attacks.
Our model is based upon Jones calculus \cite{jones42} and past experimental characterizations of the PIC PUF components \cite{macfarlane23,shahoei22}.
We validate the model by analyzing properties of the CRP distribution, including autocorrelation, bijectivity, challenge-response relationship, and quantile distribution.
A detailed account of the circuit validation process is beyond the scope of this work, and will be discussed in future work. 

\subsection{Brief Introduction to Machine Learning-Based Attacks for Conventional PUFs}
``Strong PUFs'' are defined as being impossible to clone, being impossible to characterize completely within a limited time frame, and having responses that are difficult to numerically predict for any randomly selected challenge, even if many CRPs are known \cite{ruhrmair10,ruhrmair14}.
Many PUFs that were believed to be strong have proven susceptible to ML-based attacks, in which models use CRP subsets to learn predicted responses for unseen challenges.
For example, arbiter PUFs and ring oscillator PUFs were amongst the earliest conventional IC PUFs to be classified as strong PUFs \cite{naveenkumar22,mahalat19,sahoo15}, and---for some CRP bit lengths---they have been widely subjected to ML-based attacks, including regression mappings, state vector machines, $k$-nearest neighbors, random forests, genetic algorithms, and artificial neural networks \cite{delvaux19,ge20,santikellur19,machida15,khalafalla19,aghaie21,hazari21,kumar18,saha13,ganji16,yoon20}.

Given adequate time and CRPs, ML models can often predict response bits with a probability greater than chance, increasing the susceptibility of a given PUF \cite{delvaux19,hazari21,kumar18,ge20,santikellur19,machida15,aghaie21,khalafalla19,saha13,ganji16,laguduva20,guo16}.
While subsequent adjustments or complements to PUF design may neutralize the specific ML-based attack under consideration \cite{hazari21,ge20,dodda21,machida15,laguduva20}, iterative improvements to the ML approach can then compromise such enhanced PUFs \cite{khalafalla19}.
Furthermore, theoretical results have proven both unfavorable statistical properties and bounds on the number of CRPs required to train ML-based attacks to a desired degree of accuracy, thereby illustrating that some types of PUF previously thought to be strong might not merit that classification \cite{siddhanti2019analysis,ganji16,ganji16a,ganji15,ganji16b}.
Given the pressures of ML-based attacks on PUF functionality, it is important to assess the photonic PUF's susceptibility to such attacks.

\section{Attacking the Photonic PUF}\label{sec:attacking_the_puf}

\subsection{Modeling the PUF}\label{sub:modeling_the_puf}
Our PUF model accepts a $24$-bit challenge and produces a $24$-bit response; these bitstrings are derived from decimal-value representations of the optical signal input to the PUF and the measured signal output.
MacFarlane et al. explains the structure of these challenge and response bitstrings \cite{macfarlane23}; for purposes of this paper, we emphasize that the bitstrings can be represented as binary 24-bit values or as varying-digit MVL values.

\begin{figure}[ht]
\centerline{\includegraphics[width=\linewidth]{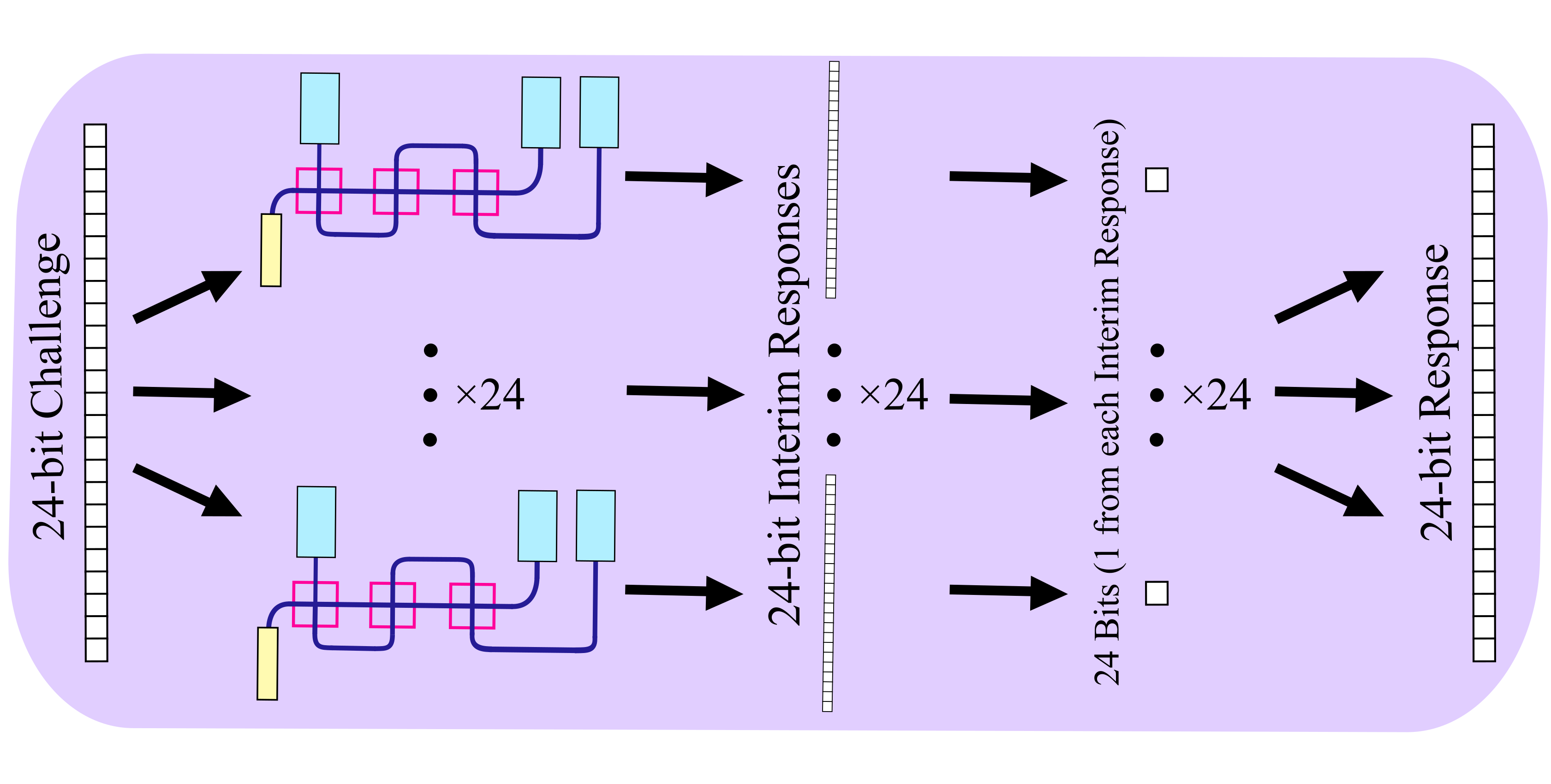}}
\caption{The $24$-cell photonic PUF architecture.}
\label{fig:puf_model_architecture}
\end{figure}

Additionally, model analysis demonstrates that we obtain a stronger PUF when we improve statistical independence of CRPs by combining several of the PUF structures of MacFarlane et al.
Specifically, as illustrated in \Cref{fig:puf_model_architecture}, we combine 24 cells, each of which is provided the same 24-bit challenge, and each of which will contribute one bit of an overall 24-bit response.  
Hereinafter, when we refer to ``the PUF,'' we are referring to the entire 24-cell architecture of \Cref{fig:puf_model_architecture}.

\subsection{The Network, Training Phase, and Evaluation Phase}\label{sub:network_training_evaluation}
\begin{figure}[th]
\centerline{\includegraphics[width=\linewidth]{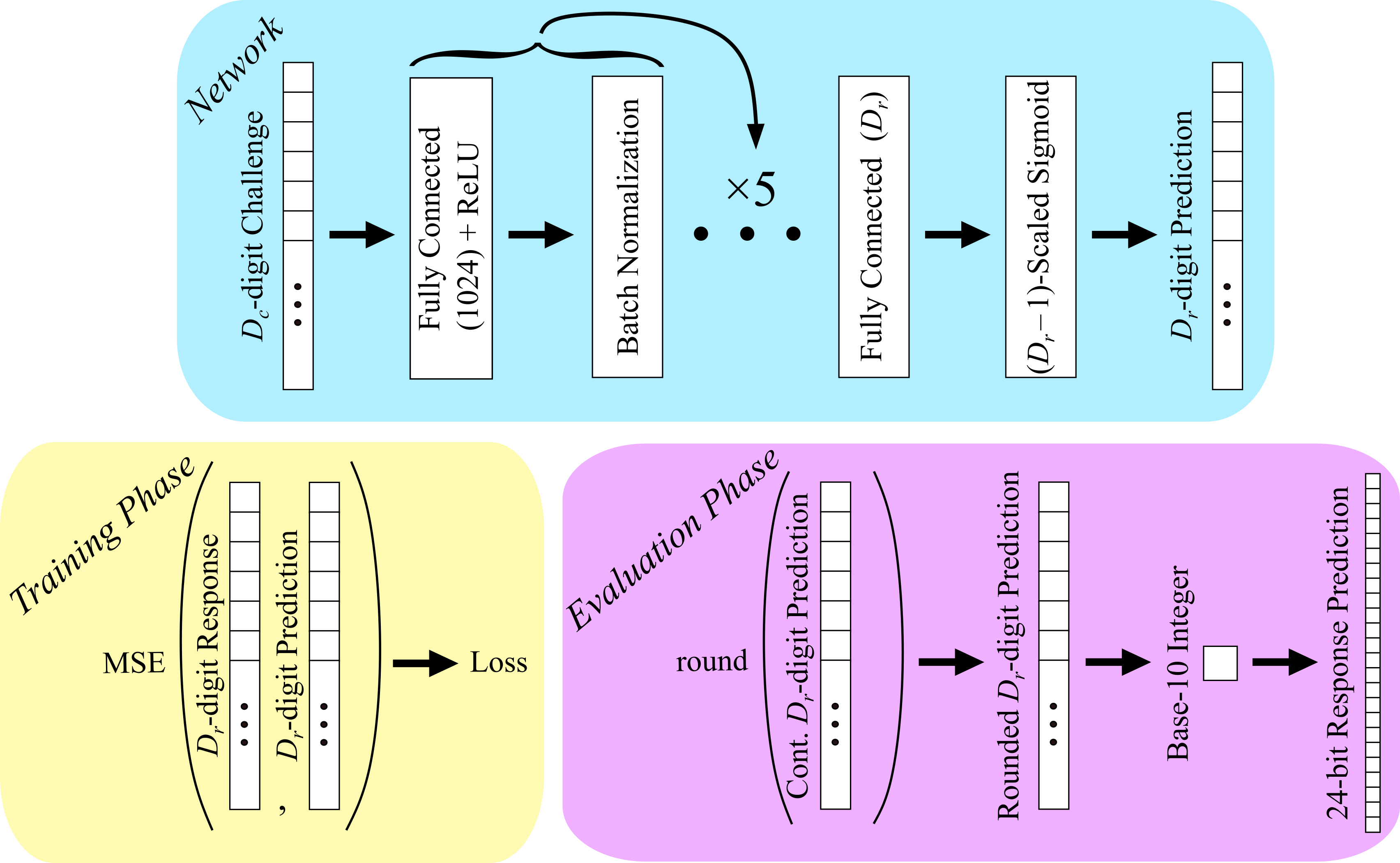}}
\caption{Overview of the neural network architecture, the training phase, and the evaluation phase. The network accepts $D_c$-digit responses encoded using radix-$R_c$ and uses six layers (described in detail in the main text) to predict the $D_r$-digit response represented using radix-$R_r$. During training, the loss is computed as the mean-squared error between the ground truth response represented as $D_r$-digit, radix-$R_r$ value and the $D_r$-digit response prediction. During evaluation, the $D_r$-digit response prediction (which includes continuous or ``Cont.'' values) is rounded.  The resulting $D_r$-digit response comprised of integer values is then converted to a base-10 integer, which is converted to a $24$-bit response vector that is compared to the ground truth $24$-bit response to assess prediction accuracy.}
\label{fig:ml_architecture}
\end{figure}

\Cref{fig:ml_architecture} provides an overview of the network architecture, training phase, and evaluation phase. We investigate several neural network models, but the purpose of each is the same: take as input a challenge vector with each element encoded using a specific radix, and predict a response vector with elements that can be interpreted using a different radix.
All architectures employ a multi-layer perceptron with six layers, which is a standard architecture for an artificial neural network.
The network's input is a $D_c$-digit, radix-$R_c$ challenge vector.
Here, $D_c$ is the number of digits required to represent the response in a radix-$R_c$ system, meaning it is given by $D_c = \lfloor \log_{R_c} N \rfloor + 1$, where $N$ is the maximum base-10 value of the response vector, or $N=2^{24} - 1$, because each challenge natively consists of $24$ bits. $D_c$-digit challenge vectors are fed into five fully-connected layers, each containing $1024$ neurons and followed by a rectified linear (ReLU) activation function \cite{nair2010rectified} and batch normalization \cite{ioffe2015batch}.

Recall that the goal is to predict responses represented with a specified radix.
So, the final layer is a fully-connected layer with $D_r$ neurons, where $D_r$ is the number of digits required to represent the response in a radix-$R_r$ system, and is computed similarly to $D_c$ above, but with $R_r$ instead of $R_c$.
To ensure output logits are in the range of $[0, D_r-1]$, model outputs are passed through a sigmoid activation scaled by $(D_r-1)$.

We use the same network architecture across all radix experiments.
However, due to variation in $D_c$ and $D_r$, the number of model parameters differs slightly for each experiment; the number of model parameters is determined by the linear layer with bias terms.
For example, when $R_c = 2$ and $R_r = 2$, there are $4$,$269$,$080$ trainable parameters.
But when $R_c = 2$ and $R_r = 16$, resulting in adjusted numbers of output neurons, there are $4$,$250$,$630$ trainable parameters.
Thus, although higher radix networks possess slightly fewer weights, the overall network complexity remains similar, ensuring a fair comparison between networks employing different MVL encodings.

During the training phase, loss is computed using the mean-squared error between the ground truth response and the predicted response, with both represented using radix-$R_r$.
Each model is trained with the Adam optimizer \cite{2015-kingma} using a cosine learning rate schedule with warm (\textit{i.e.}, not random) restarts \cite{loshchilov2017sgdr}.
The initial learning rate is set to $0.001$ with $1$,$000$ initial decay steps.
A multiplier of $2.0$ is applied to the number of steps for each subsequent restart with no amplitude attenuation applied to the learning rate.
Because we evaluate models using different CRP training set sizes, we use a batch size of $\max(\text{number of training samples}, 128)$ and a minimum of $1$,$000$ training steps to ensure models with fewer training samples are given adequate training iterations.
Training is terminated if training loss does not decrease by more than $0.0001$ following at least $100$ training steps.

During the evaluation phase we assess the ability of a given model to predict MVL responses on the CRP test set.
To fairly compare accuracy between models using different MVL response representations, we convert all response predictions to a single radix representation before computing accuracy. This also ensures that the chance prediction is 50\% for all experiments.
Specifically, we convert to the representation of the ground truth responses, which---as described in \Cref{sub:modeling_the_puf}---means converting the $D_r$-digit, radix-$R_r$ output predictions to $24$-bit ($D_r = 24,\,R_r = 2$) representations.
This process has three steps.
First, rounding: all models, regardless of radix, predict responses that have floating point---not integer---values for each of the $D_r$ digits.
Consequently, the predictions are rounded to the nearest integer, generating a $D_r$-digit response with one integer per radix-$R_r$ digit.
Second, this rounded $D_r$-digit response is converted to a base-10 integer.
Third and finally, that base-10 integer is represented as a $24$-bit response.
Accuracy is computed by averaging the binary accuracy across all $24$ bits of the predicted and ground truth responses.

\begin{figure*}[ht]
\centerline{\includegraphics[width=\textwidth]{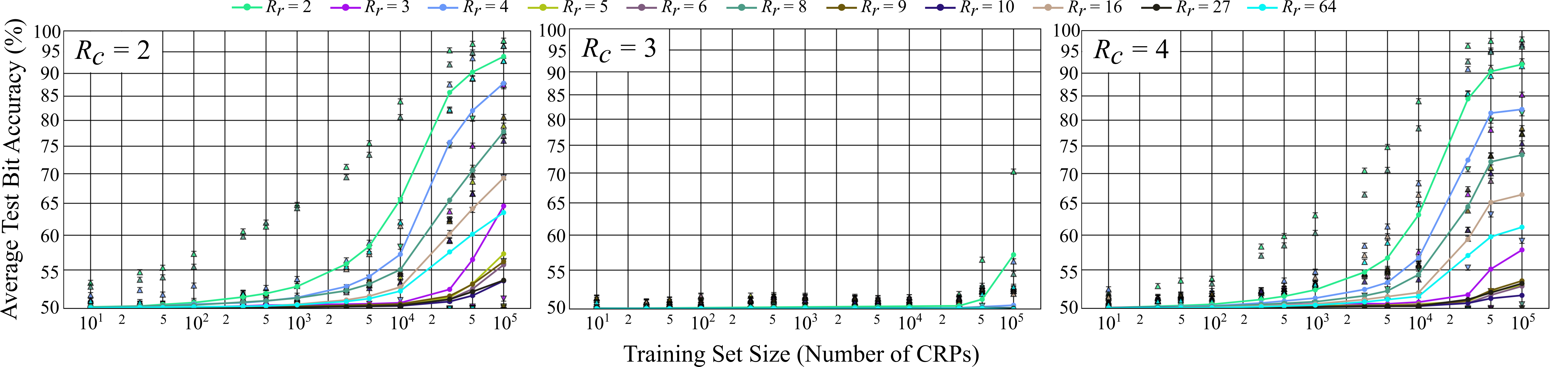}}
\caption{Response prediction accuracy for $6$,$600$ models with challenges encoded using three different radices.
Each reported accuracy is the likelihood of correctly predicting any bit in the test set averaged across validation folds and the four PUFs.
Upward-pointing triangles indicate the maximum individual bit accuracy over all PUFs and folds for the given challenge/response radix combination. Downward-pointing triangles indicate the minimum accuracy for the same. 
}
\label{fig:ml_results}
\end{figure*}

To evaluate model performance with different training set sizes, we consider results from $6$,$600$ trained models.
Specifically, we use data from four simulated PUF realizations.
For each PUF, we consider challenges represented using radices $R_c = \{2, 3, 4\}$, each of which is paired with responses represented using radices $R_r = \{2, 3, 4, 5, 6, 8, 9, 10, 16, 27, 64\}$.
These radices were chosen as a representative sample that is small enough to present and discuss, but large enough to illustrate relevant trends.
For each combination (\textit{i.e.}, $\text{PUF realization}, R_c, R_r$), we then vary training set size, considering sets of $\{10, 30, 50, 100, 300, ..., 10^5\}$ CRPs.
For each training set size, we apply $5$-fold cross-validation, each with a test set comprised of $2 \times 10^5$ CRPs disjoint from the training set.
In the next section, we describe and discuss the response prediction accuracies of these models.

\section{Results and Discussion}\label{sec:results_and_discussion}
\Cref{fig:ml_results} summarizes the accuracy of the $6$,$600$ trained models.
The results are organized by $R_c$ (the radix used to represent the challenges); specifically, the left subplot has $R_c=2$, the center subplot has $R_c=3$, and the right subplot has $R_c=4$.
Each color corresponds to results from models with a single $R_r$ value (the radix used to represent the responses).
For example, the top curve in all three subplots (in turquoise green) corresponds to $R_r = 2$ for each of $R_c = 2, 3, $ and $4$, while the bottom curve (in royal purple) on the first subplot for $R_c = 2$  corresponds to $R_r = 10$.
The points connected by lines illustrate the average accuracy of predicting any single bit of the $24$-bit response for models trained across the different PUF realizations and folds for a given $R_c$ and $R_r$.
The upward- and downward-pointing triangles illustrate the maximum and minimum individual bit accuracy achieved across PUF realizations and folds for a given $(R_c, R_r)$ pair.
It is worth emphasizing that---for all results in the plots---the accuracy presented is the likelihood of correctly predicting any bit in the test set.
For the results of \Cref{fig:ml_results}, the overall test set had $2\times 10^5$ responses (meaning $(2\times 10^5) \times  24 = 4.8\times 10^6$ bits), so a reported accuracy of $0.75$ would mean that $3.6\times 10^6$ of those bits were predicted correctly.
This accuracy metric facilitates straightforward and equitable comparison of models using different radices.
Specifically, as is discussed in \Cref{sub:network_training_evaluation}, the final $D_r$-digit radix-$R_r$ response prediction is converted to a $24$-bit representation, because the value of ``chance'' is different for each radix.
Such conversion means that, for all results in \Cref{fig:ml_results}, a $50\%$ average accuracy denotes that the model was---on average---no better than chance at predicting any given bit of the $24$-bit response, while a $50\%$ maximum accuracy means that the model was never able to predict any one of the $24$ response bits better than chance.
For example, the bright green line in the left plot corresponding to $(R_c = 2, R_r=2)$ illustrates that when using $3\times 10^3$ training CRPs, the average accuracy of predicting a response bit is approximately $55\%$, and the maximum single bit accuracy achieved by any $(R_c = 2,  R_r = 2)$ model trained on $3\times 10^3$ CRPs is approximately $71\%$.

We begin analysis with two observations regarding the number of CRPs required to train to specific prediction accuracies.
First, the accuracies increase monotonically; the average, maximum, and minimum accuracy either stays the same or increases as the number of CRPs on which the model trains increases.
Consequently, and as expected, the highest maximum, average, and minimum accuracies are associated with models trained on the largest number of CRPs tested ($10^5$).
While those accuracies can be far higher than chance (most notably $94$\% and $92$\% for the $(R_c = 2, R_r = 2)$ and $(R_c = 4, R_r = 2)$ cases, respectively), there are two noteworthy points.
First, for five of the response radices when $R_c=2$, for 
\textit{all} of the response radices when $R_c=3$, and for six of the response radices when $R_c=4$, the average accuracy remains less than $60\%$, indicating that in the majority of challenge/response radix pairs tested, the average accuracy remains less than $60\%$ even when $10^5$ CRPs are used for training.
This is significant because of the second point, which is that obtaining $10^5$ CRPs for a functional PUF is a difficult task for an attacker.
Particularly for photonic PUFs, which are harder to scrape for CRPs than their conventional counterparts, obtaining even a fraction of that many CRPs is challenging and should be indicative of a failure beyond the PUF's inherent susceptibility to MVL-based ML attacks.

Second, the plots illustrate that at least $10^3$ CRPs are required to train models that predict individual bits better than chance for all challenge and response radix combinations tested.
Furthermore, $10^3$ training CRPs often puts model performance just slightly above chance: in $31$ out of $33$ radix combinations tested, even $10^4$ CRPs is not enough to train models that predict above $60\%$ accuracy.
As with the discussion above, collecting this many accurate CRPs should be an infeasible task for a functioning PUF.

Before closing, we briefly discuss three additional---and more qualitative---observations that present interesting avenues for further study.
First, models that use lower-radix response representations generally perform better than higher-radix counterparts.
For example, the models that use $R_r=2$ perform the best across $R_c=\{2,3,4\}$, while the models with $R_r=\{5,6,9,10,27\}$ are always amongst the worst performers.
We do not have quantitative evidence indicating why this is the case.
However, we hypothesize that this may be an effect of aggregated bit relationships drowning out more specific relationships that are learned more easily when bits are kept separate, or at least combined to a lesser extent with lower-radix responses.
In other words, higher-radix encodings may obfuscate response bit relationships, presenting a more challenging task for networks to learn.

Second, both challenge and response radices that are powers of two are more easily predicted, sometimes by substantial margins.
For all challenge radices, the models trained on response radices that are powers of two are more easily predicted than nearly all other non-power of two radices.
Furthermore, the models trained with $R_c=3$ are more difficult to predict compared with $R_c$ values that are powers of two: across all response radices paired with $R_c = 3$, the maximum and average accuracies increase from chance only when trained using the largest number of considered CRPs.
Again, it is not immediately clear why this is the case.
One possible factor is the native $R_r=2$ representation---and thus the native resolution---of the CRPs.
Recall from \Cref{sub:modeling_the_puf} that the challenges and responses are recorded as $24$-digit bitstrings that are thus written using radix-$2$.
The physical constraints of measurement devices used to interact with fabricated PUFs dictate that these bitstrings take on only a portion of the possible $2^{24}$ values; it is known that resolution constraints affect what values can be recorded as CRPs for a realized PUF.
It is possible that this radix-$2$-determined resolution may inform the relationships between challenges and responses represented using radix-$2$ in ways that are dulled---or entirely lost---in higher radices that are not powers of two.

This low model performance when CRPs are represented with radix-$3$ might provide a straightforward avenue to stronger photonic PUFs: models trained using CRPs natively represented with radix-$3$ should be unable to accurately predict response bits.
Of course, such representation would not be an insurmountable obstacle, because an adversary who both knew of differing radix performance and who had enough CRPs to guess that radix-$3$ was used could convert to radix-$2$ before training.
Nonetheless, radix-$3$ CRP representation would thus require additional knowledge and effort from would-be attackers, thereby further lessening the photonic PUF's susceptibility to these ML attacks.

Third and finally, the maximum accuracy attained over all models also varies with $R_c$ and $R_r$, and the trend tracks that of the average accuracy: radices that are powers of two provide higher maximum accuracies with smaller training sets.
Additionally, the maximum accuracies increase more rapidly than the averages: in the $R_c = 2$ case, the maximum accuracy is above chance with a training set containing two orders of magnitude fewer CRPs than is required to achieve an above-chance average accuracy (on the order of $10^1$ versus on the order of $10^3$).
A similar result holds for the $R_c = 4$ case, where maximum accuracies above chance require training sets with approximately $10^2$ CRPs, and not the approximately $10^3$ required for above-chance average accuracies.
Furthermore, with the $R_c = 2$ and $R_c = 4$ cases, maximum accuracies can be well above chance, up to $98.5$\%, when $10^5$ CRPs are included in the training set.
These results do not necessarily pose immediate risks to the photonic PUF's resilience because the PUF has a $24$-bit response.
As is illustrated by the average accuracies that---for all feasible CRP training set sizes---remain substantially lower than above-chance maximum accuracies, there are enough bits that are not easily predictable so as to still provide PUF resilience to attacks with fewer than $10^3$ CRPs in the training set.
Nonetheless, the maximum accuracies are noteworthy because they indicate a possible weakness that might be exploitable with further study.
Specifically, although $10^3$ CRPs should still be beyond tractable attainment, it is possible that an attack might leverage a less-resilient bit that can be predicted with an average accuracy of, say, $61$\% with $300$ CRPs (as is true of the $(R_c = 2, R_r = 2)$ case) to attain information about other generally-less-predictable bits.
This possibility makes better understanding which bits are most predictable---and why---an important avenue for further research.
 
\section{Conclusion and Future Work}
Increasingly sophisticated ML-based attacks require study of the susceptibility of both well-established and novel PUFs.
We assess the susceptibility of the photonic PUF introduced in MacFarlane et al. to MVL-based ML attacks and---based upon data from a simulated model of the PUF---we conclude that it exhibits resilience to these attacks.
Specifically, when fewer than $1$,$000$ CRPs are available for training, models of any radix considered (between $2$ and $64$) cannot---on average---predict responses better than chance, meaning each bit of the response can be predicted with only 50\% accuracy.
Furthermore, while training sets of approximately $1$,$000$ CRPs can attain accuracies better than chance for some radices, it should be very difficult for an attacker to obtain that many accurate CRPs with which to train.
Consequently, our results suggest that---in practically-feasible cases---ML-based MVL attacks would likely not be successful in predicting responses corresponding to as-of-yet-unseen challenges for the photonic PUF.
Nonetheless, there is substantial future work, including study of the different MVL model and bit performance, of longer PUF responses such as those used in cryptographically-secure applications, and of the PUF's statistical properties as demonstrated by the model, including uniqueness and uniformity.
Such analysis could inform adjustments to PUF structure---including our approach of combining PUF cells---that might improve resilience beyond the approximately $1$,$000$ CRP training set size.

\section*{Acknowledgment}
The authors gratefully acknowledge funding from Anametric, Inc.

\bibliographystyle{IEEEtran}
\bibliography{references}

\end{document}